\documentclass[nocompress]{spie}

\usepackage{amsmath,amsfonts,amssymb}
\usepackage{graphicx}
\usepackage{xcolor}
\usepackage{subcaption, floatrow, arydshln}
\usepackage[colorlinks=true, allcolors=blue]{hyperref}
\usepackage[normalem]{ulem}
\usepackage{color, soul}

\title{Design and implementation of a noise temperature measurement system for the Hydrogen Intensity and Real-time Analysis eXperiment (HIRAX)}
\author[a]{Emily R. Kuhn}
\author[a, b]{Benjamin R.~B. Saliwanchik}
\author[a]{Maile Harris}
\author[c]{Moumita Aich}
\author[d]{Kevin Bandura}
\author[e]{Tzu-Ching Chang}
\author[c, f]{H. Cynthia Chiang}
\author[c, g]{Devin Crichton}
\author[h]{Aaron Ewall-Wice}
\author[c]{Austin A. Gumba}
\author[i]{N. Gupta}
\author[c]{Kabelo Calvin Kesebonye}
\author[j]{Jean-Paul Kneib}
\author[k]{Martin Kunz}
\author[c, l]{Kavilan Moodley}
\author[a]{Laura B. Newburgh}
\author[k]{Viraj Nistane}
\author[c]{Warren Naidoo}
\author[f]{Deniz \"{O}l\c{c}ek}
\author[m]{Jeffrey B. Peterson}
\author[g]{Alexandre Refregier}
\author[f, n]{Jonathan L. Sievers}
\author[o]{Corrie Ungerer}
\author[k]{Alireza Vafaei Sadr}
\author[p]{Jacques van Dyk}
\author[q]{Amanda Weltman}
\author[f]{Dallas Wulf}

\affil[a]{Department of Physics, Yale University, New Haven, CT, USA}
\affil[b]{Department of Physics, Brookhaven National Laboratory, Upton, NY, USA}
\affil[c]{School of Mathematics, Statistics \& Computer Science, University of KwaZulu-Natal, Westville Campus, Durban4041, South Africa}
\affil[d]{Department of Computer Science and Electrical Engineering, and Center for Gravitational Waves and Cosmology, West Virginia University, Morgantown, WV, USA}
\affil[e]{Jet Propulsion Laboratory, California Institute of Technology, Pasadena, CA, USA}
\affil[f]{Department of Physics, McGill University, Montr\'eal, QC, Canada}
\affil[g]{Institute for Particle Physics and Astrophysics, ETH Z\"{u}rich, Z\"{u}rich, Switzerland}
\affil[h]{Department of Astronomy and Physics, UC Berkeley, CA, USA}
\affil[i]{Inter-University Centre for Astronomy and Astrophysics, Post Bag 4, Ganeshkhind, Pune 411 007, India}
\affil[j]{Institute of Physics, Laboratory of Astrophysics, Ecole Polytechnique F\'ed\'erale de Lausanne (EPFL), Observatoire de Sauverny, 1290 Versoix, Switzerland}
\affil[k]{D\'epartement de Physique Th\'eorique and Center for Astroparticle Physics, University of Geneva}
\affil[l]{Astrophysics Research Centre, University of KwaZulu-Natal, Westville Campus, Durban 4041, South Africa}
\affil[m]{Department of Physics, Carnegie Mellon University. Pittsburgh. PA, USA}
\affil[n]{School of Chemistry and Physics, University of KwaZulu-Natal, Durban, South Africa}
\affil[o]{ArioGenix(Pty) Ltd, Pretoria, South Africa}
\affil[p]{Pronex Engineering Management Consultants CC, Pretoria, South Africa}
\affil[q]{Department of Mathematics and Applied Mathematics, University of Cape Town, South Africa}

\authorinfo{Further author information: (Send correspondence to E.R.K.)\\E-mail: emily.kuhn@yale.edu\\}

\begin{document} 
\maketitle

\begin{abstract} 
This paper describes the design, implementation, and verification of a test-bed for determining the noise temperature of radio antennas operating between 400-800\,MHz. The requirements for this test-bed were driven by the HIRAX experiment, which uses antennas with embedded amplification, making system noise characterization difficult in the laboratory. The test-bed consists of two large cylindrical cavities, each containing radio-frequency (RF) absorber held at different temperatures (300\,K and 77\,K), allowing a measurement of system noise temperature through the well-known `Y-factor' method. The apparatus has been constructed at Yale, and over the course of the past year has undergone detailed verification measurements. To date, three preliminary noise temperature measurement sets have been conducted using the system, putting us on track to make the first noise temperature measurements of the HIRAX feed and perform the first analysis of feed repeatability.
\end{abstract}


\section{INTRODUCTION}
\label{sec:intro} 
\begin{figure}[ht]
\begin{center}
\includegraphics[width=\textwidth]{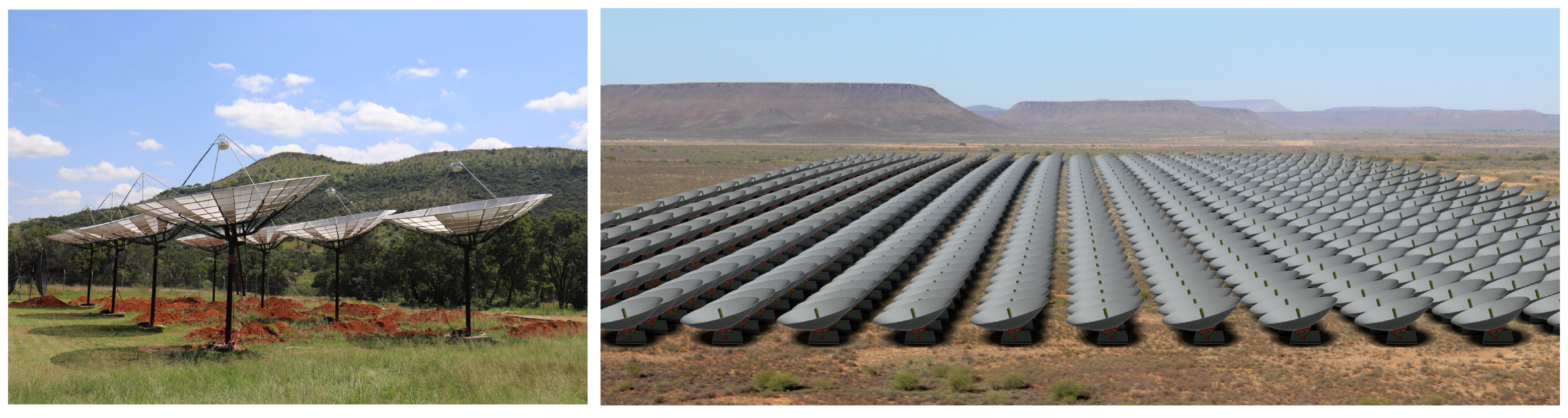}
\end{center}
\caption{The HIRAX prototype array at Hartebeesthoek Radio Astronomy Observatory, South Africa (left) and a rendering of the final 1024 dish configuration with the current prototype dish model in the Karoo Desert, South Africa (right).}
\label{hirax}
\end{figure}

The Hydrogen Intensity and Real-time Analysis eXperiment (HIRAX) is a 21\,cm neutral hydrogen intensity mapping experiment to be deployed in the Karoo Desert in South Africa~\cite{hirax}. It will consist of 1024 six-meter parabolic dishes \cite{benpaper}, and will map much of the southern sky over the course of four years. HIRAX is designed to improve constraints on the dark energy equation of state through measurements of large scale structure at high redshift. It will target a measurement of the $100 h^{-1}$Mpc Baryon Acoustic Oscillation scale through 21\,cm emission of neutral hydrogen contained in abundance in galaxies. The HIRAX redshift range, 0.8 $< z < 2.5$, corresponds to the radio band of 400-800\,MHz and will be measured in 1024 frequency bins. In addition to 21cm cosmology and hydrogen absorber science, HIRAX will discover and monitor transients such as fast radio bursts (FRBs) and pulsars, as is currently done with CHIME in the Northern Hemisphere~\cite{frbboyle2018first,frbamiri2019second,frbamiri2019observations,frbandersen2019chime,frbamiri2020periodic,frbandersen2020bright}. HIRAX's southern location will allow for a variety of cross-correlation measurements with other cosmology surveys such as ACTPol, DES, and the Vera Rubin Observatory. Currently, an 8-element prototype array has been deployed at Hartebeesthoek Radio Astronomy Observatory (HartRAO), and a 256-element array is being developed at the final HIRAX site (Figure \ref{hirax}).

\begin{figure}[ht]
\begin{center}
\includegraphics[width=0.3\textwidth]{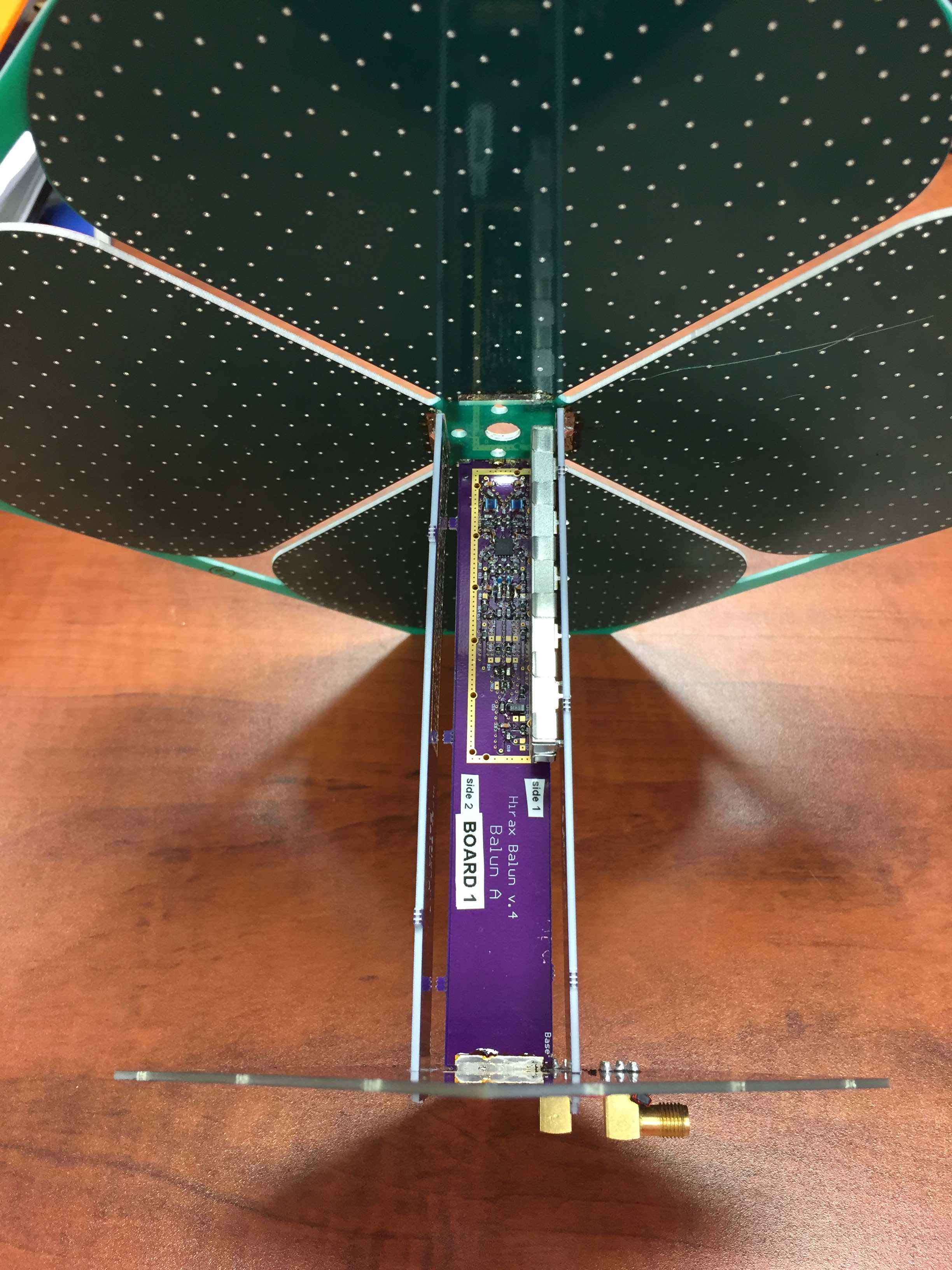}
\end{center}
\caption{The HIRAX antenna. The HIRAX antenna feed has the first stage amplification integrated into the antenna structure. This reduces feed noise and makes it more lightweight, but in turn prevents the noise temperature from being directly measured with a noise figure meter.}
\label{balun}
\end{figure}

The HIRAX signal chain is comprised of both custom and commercial parts\cite{hirax}. Each of the 1024 dishes will have a dual-polarization antenna feed with a first-stage differential low-noise amplifier (LNA, Avago MGA--16116) embedded directly into the antenna balun for noise reduction (Figure~\ref{balun}). This design choice has been adopted to keep the total system noise to less than 50\,K, allowing a sensitive measurement of the $\sim$100$\mu$K cosmological 21\,cm signal\cite{shaw2015coaxing}. The signal is transformed to an optical signal using an RF-over-Fiber (RFoF) module\cite{rfof}, and then carried to the correlator building, where it is turned back to RF. The analog signal is filtered and further amplified, before it is channelized and digitized by an ICE board\cite{bandura2016ice}, and then correlated.

The HIRAX feed is a dual polarization cloverleaf dipole with low loss ($<$0.15\,dB) and small reflectivity ($<-15$\,dB). It consists of FR-4 dielectric (PCB) with a metalized layer, a PCB balun and support board. A ring choke is used to circularize the beam and decrease crosstalk and ground spillover. The cloverleaf design is based on that of the CHIME antenna \cite{deng14}, with the main differences being HIRAX uses FR-4 instead of Teflon-based PCB (to reduce cost) and sums polarization along a different axis. 

\begin{table}[h!]
\centering
 \begin{tabular}{|l|l|} 
 \hline
Frequency Range & 400--800\,MHz\\ 
 \hline
Frequency Resolution & 390\,kHz, 1024 channels\\
 \hline
Dish size & 6\,m diameter, f/D = 0.23\\
 \hline
Field of View & 15-56\,deg$^2$\\
 \hline
System Temperature & $\sim$50 Kelvin\\
 \hline
Antenna Noise Temperature & 20-30 Kelvin\\
\hline
 \end{tabular}
 \caption{ \label{tab:hiraxspecs}
HIRAX instrument parameters.}
\end{table}

The focus of this work is on assessing the HIRAX feed contribution to system noise. To measure cosmological emission efficiently we require the total noise to be kept below 50K, of which at most 30K can come from the feed itself. The feed design is optimized to reduce system noise by removing particular sources of loss in the analog chain, primarily by moving the first stage amplification into the antenna balun. The choice to embed amplification directly into the balun stem makes noise temperature characterization challenging, as the noise temperature, gain, and stability of the amplifier cannot be measured directly with a noise figure meter or related laboratory equipment. To measure the noise temperature of the feed and amplifier, we must inject a broad-band signal into the HIRAX antenna. 
In this paper, we describe a test-bed built at Yale University that allows us to measure signals from radio absorber at two known temperatures, and infer the noise temperature of the antenna and amplifier via the Y-factor method (see e.g. \textit{Microwave Engineering} \cite{pozar}).

The Y-factor method allows one to determine the noise temperature of an antenna by comparing the output power from two loads at different temperatures. This method can be derived from the observation that an antenna in a cavity in thermal equilibrium at temperature T behaves like a resistor of temperature T, where the antenna internal noise can be likened to that of Johnson noise in the resistor\cite{dicke}. It is most commonly expressed through the following set of equations\cite{pozar}: 

\begin{equation}
Y=\frac{P_{\text{hot}}}{P_{\text{cold}}}
\end{equation}

\begin{equation}
T_{\text{noise}}=\frac{T_{\text{hot}}-YT_{\text{cold}}}{Y-1},
\end{equation}
where $P_{\text{hot}}$ is the measured power from a ``hot'' load, $P_{\text{cold}}$ is the measured power from a ``cold'' load, and $T_{\text{hot}}$, $T_{\text{cold}}$ are the corresponding load temperatures. $Y$ is referred to as the ``Y Factor''. This measurement requires linearity in the gain of active components and known sources of input power.

This type of measurement is commonly taken using the sky as a cold load, but this is impractical for the HIRAX feed, which has a wide beam that is still being characterized, and for us locally due to the radio frequency interference (RFI) rich environments around universities. We instead designed and constructed an experimental system with hot/cold loads of 300K/77K (corresponding to room temperature and liquid nitrogen temperature) to be used in a laboratory setting. We aim to understand the antenna noise temperature to within 5K, or 10\% of the expected system noise.

\section{Experiment Design}
The noise temperature measurement system has been designed as a pair of reflective closed cylindrical cavities with radio-frequency (RF) absorber in the bottom. One cavity is kept at ambient temperatures such that the absorber emits at $T_{\rm hot}$ = 300\,K, and designated the `warm' load. The second cavity is filled with liquid nitrogen such that the absorber emits at $T_{\rm cold}$ = 77\,K, and is designated the `cold' load. The feed is attached to the lid of each cavity such that its beam terminates in the absorber. There were several constraints on the design, including:

\begin{itemize}
\item Cost: We use commercially available materials with minimal labor to assemble. 
\item Size: We required that the cavities are smaller than $\sim$1.5\,m, to stay within a reasonable footprint in the lab space, and ensure standard material stock is suitable for construction. 
\item Cavity material: We use steel because it is easily weldable (important for containing liquid nitrogen as a safeguard) and inexpensive relative to other materials.
\item Absorber: We use commercially available AEP-18 series pyramidal foam\footnote{https://www.mvg-world.com/en/products/absorbers/standard-absorbers/pyramidal-absorbers-aep-series} for the RF absorber, which has $\sim$30\,dB absorption in our band.
\item Shape: We optimize the cavity dimensions to minimize resonances, and to be structurally stable and capable of containing 600\,L of liquid nitrogen. 
\item Insulation: We line all surfaces of the chamber with insulation to reduce liquid nitrogen boil-off and limit the accumulation of water vapor. We also add a foam lid to encourage a nitrogen vapor layer and isolate the feed from the cold vapor. 
\item Reflectivity: We coat the inside of the cavities with aluminum tape to increase the reflectivity, primarily to improve the hold-time of the liquid nitrogen in the cavity. 
\item Indistinguishable: The two cavities must be sufficiently similar, or their differences sufficiently repeatable and characterized, to keep systematic errors less than 5\,K. Cavity differences will be quantified in more detail in later sections.  
\end{itemize}

As described in this section, the design of the test-bed was optimized using CST simulations prior to construction, and materials were chosen to allow one of the cavities to hold liquid nitrogen.

\subsection{Simulations}

We optimized the design of the cavities using CST Microwave Studio\footnote{https://www.3ds.com/products-services/simulia/products/cst-studio-suite/solvers/}. CST was a natural choice for modeling, as the HIRAX and CHIME collaborations had previously used it to construct feed models, which were leveraged for this work. We began by simulating the HIRAX feed in free space and monitoring the $S_{11}$ (return loss). Because the cavity should be an absorptive blackbody, the primary figure of merit was to match the cavity reflections, parameterized by $S_{11}$, to those of the feed in free space (such that dominant reflections are internal to the feed itself). The simulations were performed and optimized with a passive HIRAX feed attached to the lid of the cavity, with the lid functioning as a ground plane. HIRAX plans to employ a circular choke to circularize the beam, which was not included in these simulations.

The company from which we obtained RF absorber did not supply relevant material parameters, such as dielectric constant ($\epsilon$) and loss tangent ($\tan \delta$), so we created a user-defined absorptive material through the CST optimization function with guidance from literature\cite{lamb}. This optimization was performed by placing a slab of RF absorber in front of the feed in software, and sweeping material parameters until $S_{11}$ was minimized. We ultimately simulated the RF absorber as 18\,in tall pyramidal cones (numbering 16 to a 2\,ft\,$\times$\,2\,ft block, to match dimensions of those available commercially) with density 159\,kg/m${}^3$, $\epsilon$=2.7 and $\tan{\delta}$=1.

\begin{figure}[ht]
\begin{center}
\includegraphics[width=\textwidth,height=\textheight,keepaspectratio]{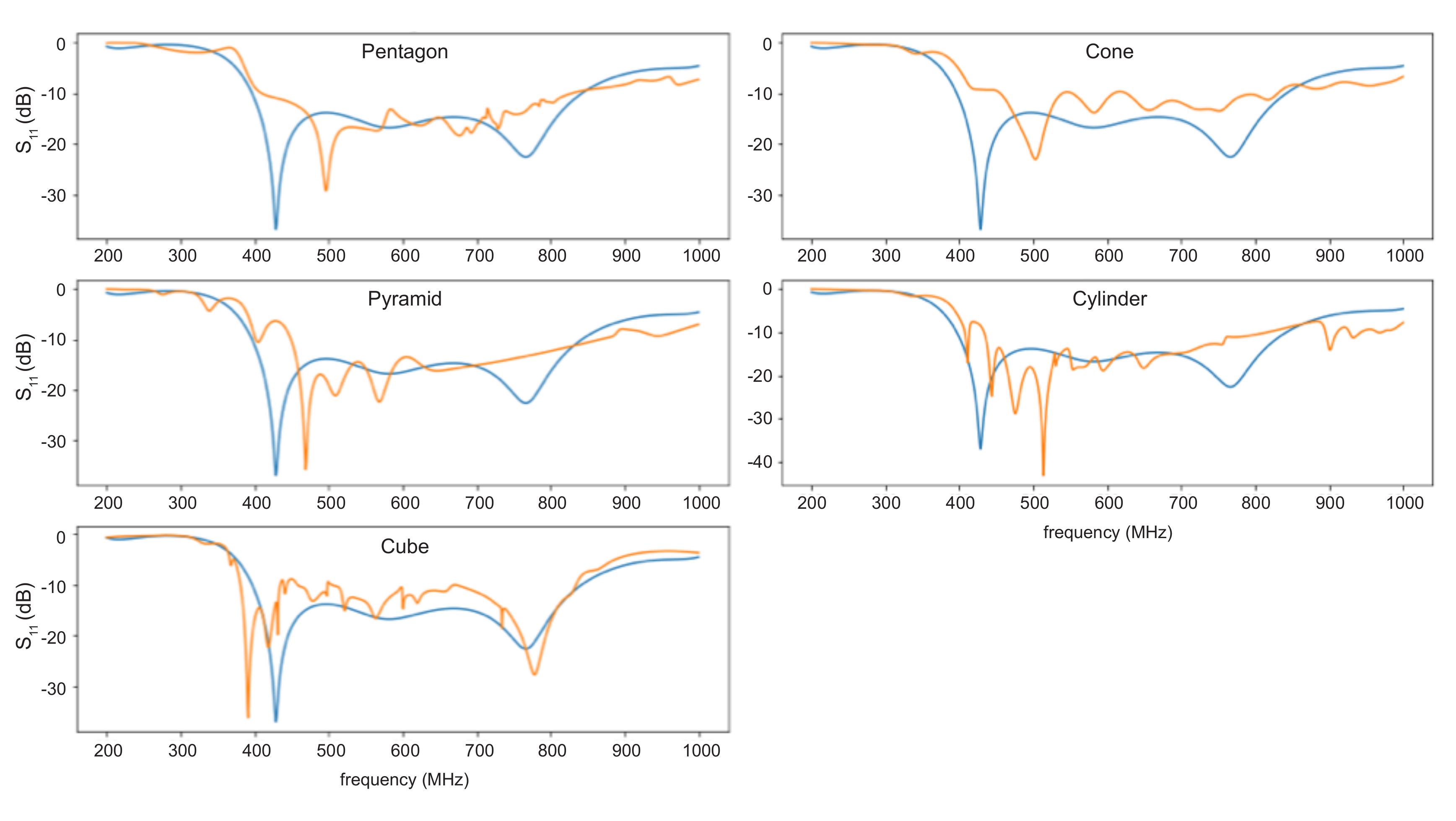}
\end{center}
\caption{Preliminary $S_{11}$ simulation results for the unamplified HIRAX feed in various cavity shapes before optimization (orange), plotted against simulation results for free space (blue). All cavities are the same height, use the same RF absorber material parameters, are made from aluminum, and share similar base dimensions (i.e.~the diameter of the cylinder and cone bases and the length of the cube sides are equal in length). From these early simulations, we find little difference in RF performance between leading contenders, and settled on a cylinder to optimize for the final design. Aside from RF performance, cylinders are mechanically robust, and simple to construct.}
\label{boxshapes}
\end{figure}

Once the RF absorber material was modeled, we set it within several steel cavities of various shapes to determine which to pursue for the design. This determination was made by comparing the feed $S_{11}$ profile of different cavity options with the free space profile. We ultimately settled on a cylinder, as it matched free space as well as the other options, is known to be the most mechanically robust shape, and is simplest to construct (Figure \ref{boxshapes}). We then optimized cavity dimensions, finding a diameter of 129.5\,cm and a height of 70.8\,cm.

We similarly estimated the reflection off of the liquid nitrogen surface by specifying $\epsilon=1.44$ and $\tan \delta=5\times10^{-5}$, which are measured parameters for 18-26\,GHz\cite{ln2_1,ln2_2} (parameters were not available in our band). We determined reflections to be below -14\,dB. As described in more detail in Section~\ref{sec:systematics}, the nitrogen is sufficiently transparent to continue with the design, but may have to be estimated and accounted for during later analysis.

Finally, we investigated tolerances on cavity dimensions by sweeping through a variety of length parameters centered about the optimal dimensions and solving for $S_{11}$. The resulting deviations in $S_{11}$ would represent possible differences between the two cavities, which we required to be less than 1\,dB. The results indicated that the cavity dimensions must be constructed with tolerance of $\sim$1\,cm. We followed a similar procedure to set a tolerance for the insulation thickness, accomplished by assuming total RF transparency of insulation, and leaving an air gap of corresponding volume in simulation. This procedure yielded an insulating later of thickness up to 80\,mm.

\begin{figure}[ht]
\begin{center}
\includegraphics[width=0.6\textwidth]{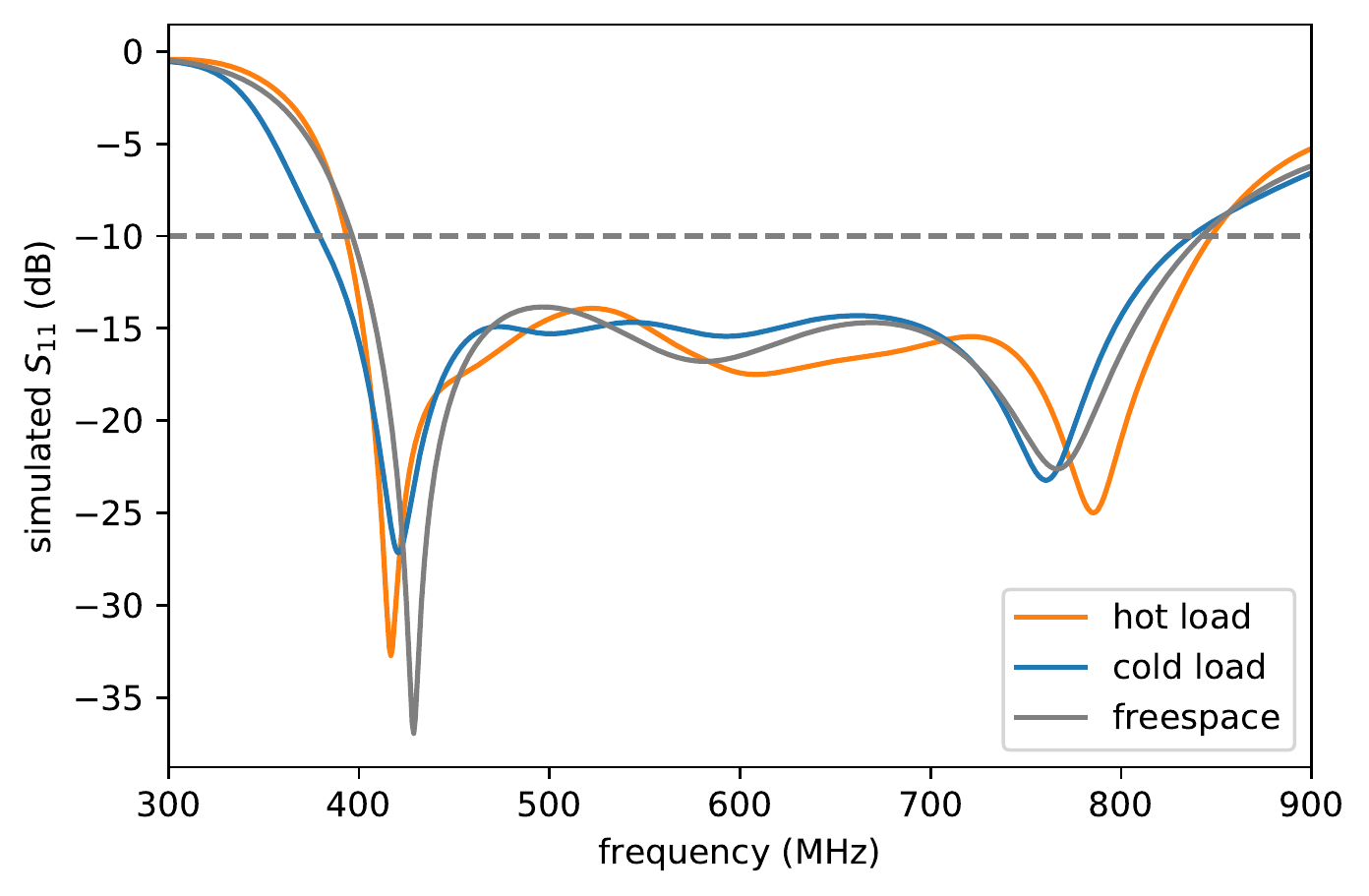}
\end{center}
\caption{$S_{11}$ simulation results for the unamplified HIRAX feed. We compare free space results to results from the Y-factor measurement system hot and cold loads (the cold load contains a simulated liquid nitrogen surface), finding all profiles to be similar to well below -10dB. Their differences are computed to give a sub-1K error in noise temperature from 400-750\,MHz.} 
\label{simulateds11}
\end{figure}

The simulated $S_{11}$ for the final, optimized cavity design is shown in Figure~\ref{simulateds11}. The resulting $S_{11}$ is similar to the free-space reflection, and below the -10\,dB reflection requirement for HIRAX. Also shown is the simulation result for reflection including the liquid nitrogen surface, which modestly changes the $S_{11}$ but remains well below -10\,dB within the 400-800\,MHz band.

\subsection{System Construction}
\label{sec:construction}
\begin{figure}[ht]
\begin{center}
\includegraphics[width=0.8\textwidth]{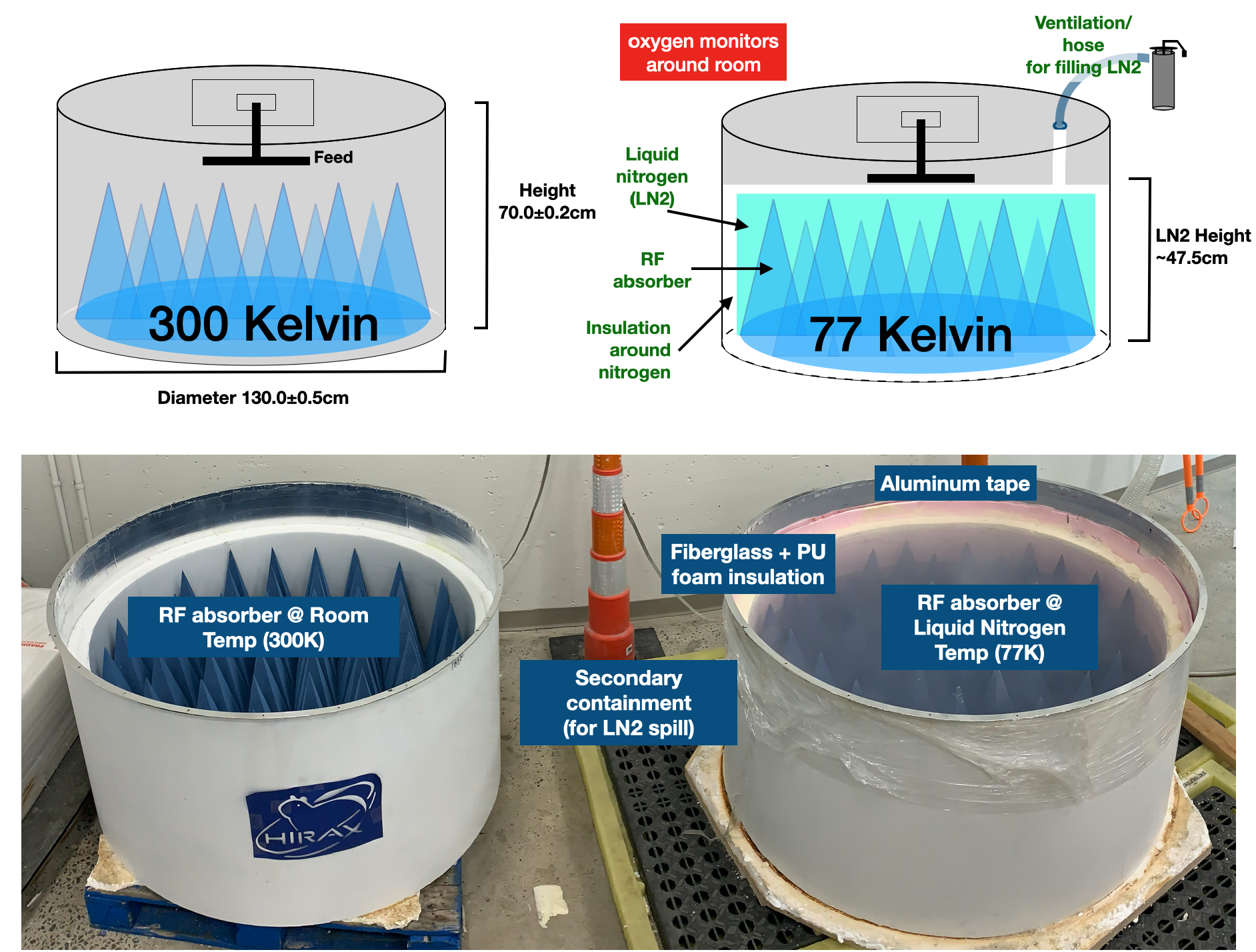}
\end{center}
\caption{(Top:) Schematic of the noise temperature measurement system design. (Bottom:) Labeled photograph of the hot/cold loads for noise temperature tests, taken during a liquid nitrogen fill. The loads are covered and sealed with a steel lid for measurements.}
\label{fig:schematicandphoto}
\end{figure}

From the simulation work described above, we determined that cylinders, 129.5 $\pm 1$\,cm in diameter and 70.8 $\pm 1$\,cm in height, were the optimal cavity shape for the noise temperature test-bed loads. The cylinders were constructed in 2018 by Welding Works in Madison, CT\footnote{http://weldingworks.com/}, formed of 1/32\,in steel with welded seams and circular removable lids, which attach to the rim of the cavities with 24 threaded screws along the circumference (see Figure \ref{fig:schematicandphoto} for full schematic). The dimensions of the cylinders were measured upon delivery, and found to be within specifications. Each lid was reinforced with two horizontal L-bars to prevent sagging. A square hole 30\,cm$\times$30\,cm was removed from the lid, so that a feed mounted to a 32.4\,cm$\times$32.4\,cm plate could be easily moved from one load to the other during measurements. The outside of the steel cylinders was painted white to prevent rusting and to decrease the radiative power load during a nitrogen fill, thus reducing the nitrogen boil off rate. Aluminium tape was applied to the cylinder interiors to decrease the emissivity further and improve nitrogen boil-off by a factor of $\sim$3.5 for an expected boil off rate of 8.6\,L/hour. 

This design required constructing an insulating cylinder capable of holding $>$550\,L of liquid nitrogen, to be inserted into one of the cavities. This insert must be radio transparent in our frequency range, closed-cell or otherwise leak-proof, and thermally insulating. We constructed small-scale test inserts of a variety of materials, including heat-sealed HD30 Zotefoam\footnote{https://www.zotefoams.com/} and conformable polyurethane spray insulations. Ultimately, a layered approach was found to meet the requirements above. The insulation consists of three layers: a fiberglass inner layer bonded to 0.6\,cm thick foamular\footnote{https://www.homedepot.com/p/Owens-Corning-FOAMULAR-1-4-in-x-4-ft-x-50-ft-R-1-Fanfold-Rigid-Foam-Board-Insulation-Sheathing-21UM/100320301}, a middle layer of 5\,cm thick cryogenically-rated Polyurethane foam \footnote{https://www.rhhfoamsystems.com/products/all-products/high-density-spray-foam/\#eluid099ab007}, and an outer layer of 0.6\,cm foamular. The fiberglass layer was constructed by placing strips of fiberglass cloth and epoxy over a mold that was then set in a vacuum-bag to remove air pockets. This fiberglass layer was placed in the ``cold'' cylinder, the cylinder walls were lined with foamular, and the cryogenically-rated Polyurethane spray foam was used to fill in the interior. The final insert is 49.5\,cm deep, with an internal diameter of 116.8\,cm and thickness of 5.1\,cm. A zotefoam lid is secured to the insulation insert to help thermally isolate the HIRAX feed from the cold nitrogen vapor, and keep the vapor shield intact during measurements.

The warm cylinder is not required to hold liquid nitrogen, so a simpler Zotefoam insert of the same internal diameter as the cold insulation was constructed to hold the RF absorber, matching the absorber placement between the two loads. As described in Section~\ref{sec:verification}, the two cylinders were measured to have $S_{11}$ and radiated power properties that are sufficiently similar to allow a measurement within the 5\,K specified error range.

For the blackbody, we used commercial 18\,in RF absorber. It is packaged in 24\,in$\times$24\,in blocks, with cones 18\,in high. We cut the RF absorber into the insert curvature using a hand-constructed hot wire foam cutter, and used a laser cut jig of appropriate curvature to guide the cutting process.

A schematic of the two cylinders is shown in Figure~\ref{fig:schematicandphoto} (upper panel), and a photo of the constructed cylinders is shown in Figure~\ref{fig:schematicandphoto} (lower panel). We have filled the cold cylinder on five occasions with $\sim$550\,L of liquid nitrogen for a total of $\sim$4 weeks of measurement time, and it has maintained its structural integrity. It takes three 230\,L nitrogen dewars to fill the cavity insert, and one dewar per day for refilling. The boil-off rate is $\sim$5.5\,L/hour, with slight variations depending on seasonal climate conditions. 

\section{System Verification Measurements}
\label{sec:verification}
A variety of measurements were performed to characterize experimental uncertainties and verify that the cavities met specifications. As described below, these measurements include verifying the radio-frequency transparency of the foam materials, quantifying the degree of similarity between the two cavities, and assessing the RF spectrum of the absorbers in each cavity. The verification measurements were primarily performed with an unamplified HIRAX feed to allow $S_{11}$ (return loss) measurements, which are not possible with the amplified HIRAX feed.

\subsection{RF transparency tests of insulation materials}

As described in Section~\ref{sec:construction}, several layers of insulation were added into the experimental system to successfully contain the 550\,L of liquid nitrogen required for this experiment. These layers must be RF transparent, as any absorption in the insulating foam will add an unquantified warm temperature component to the cold temperature measurement and bias the calculated noise temperature. To assess material transparency and inform the final design, we took $S_{11}$ measurements\footnote{A R\&S FSH4 multi-purpose analyzer was used for these measurements, www.rohde-schwarz.com/us/product/fsh-productstartpage\_63493-8180.html} of the cavity at different stages of insulation construction, which occurred over the span of several months in 2019. These measurements were performed without RF absorber such that the cavity should be highly reflective, forming a sensitive measurement of absorption in the insulating foam.

\begin{figure}[h!]
\begin{center}
\includegraphics[width=0.8 \textwidth]{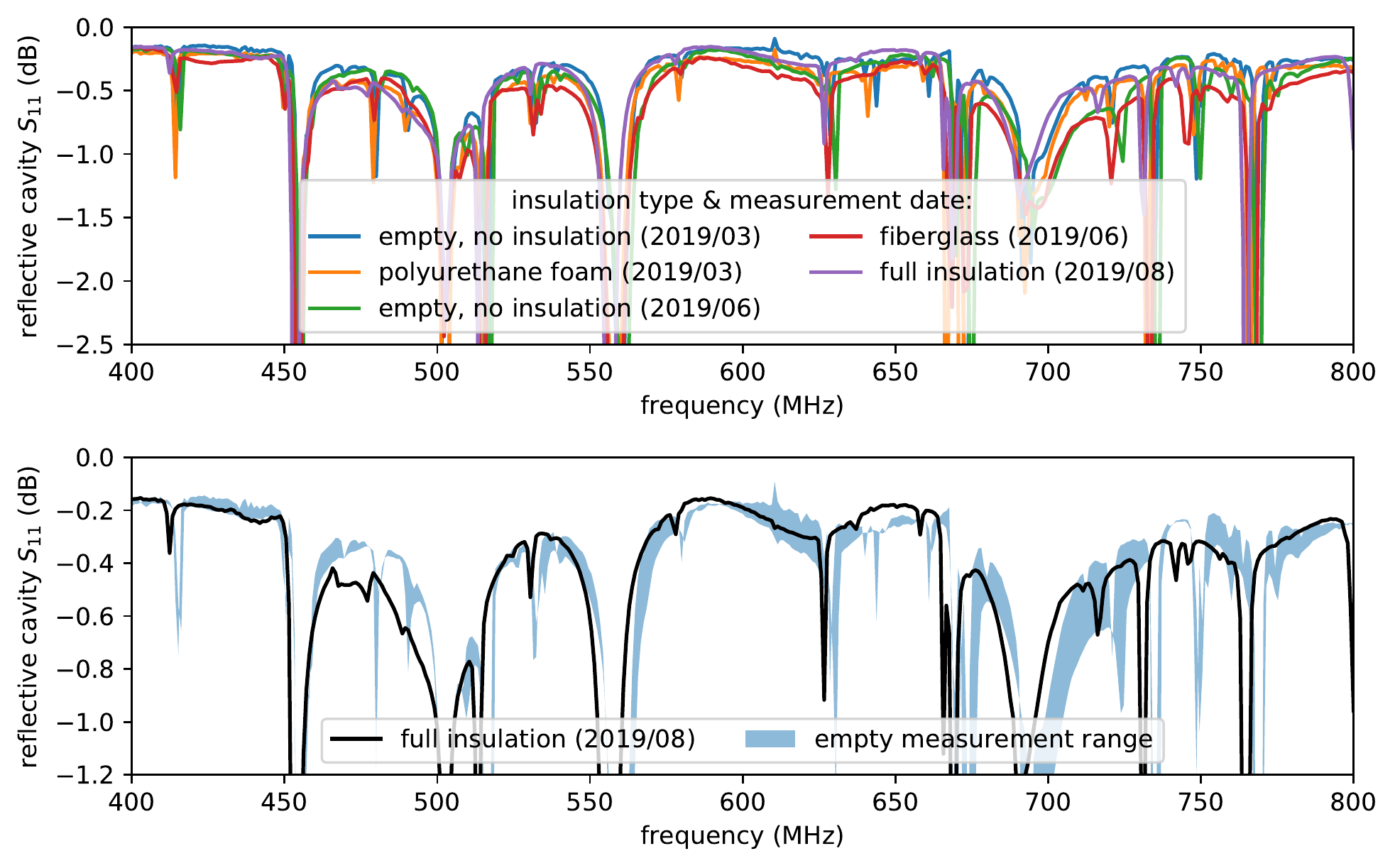}
\end{center}
\caption{(Top) Return loss ($S_{11}$) measurements of the unamplified HIRAX feed in a reflective cavity as various insulation components are added in the construction. (Bottom) Return loss measurements of the cavity with full insulation compared with the range of empty measurements. The empty cavity was measured twice (2019/03 and 2019/06), with the $S_{11}$ spectra changing by up to 0.2\,dB. The insulation components are shown to be RF transparent to within this range of fluctuations for most of the band. Not shown is the addition of an aluminum tape layer, which also has a negligible impact.} 
\label{materialtransparency}
\end{figure}

\begin{figure}[h!]
\begin{center}
\includegraphics[width=0.6\textwidth]{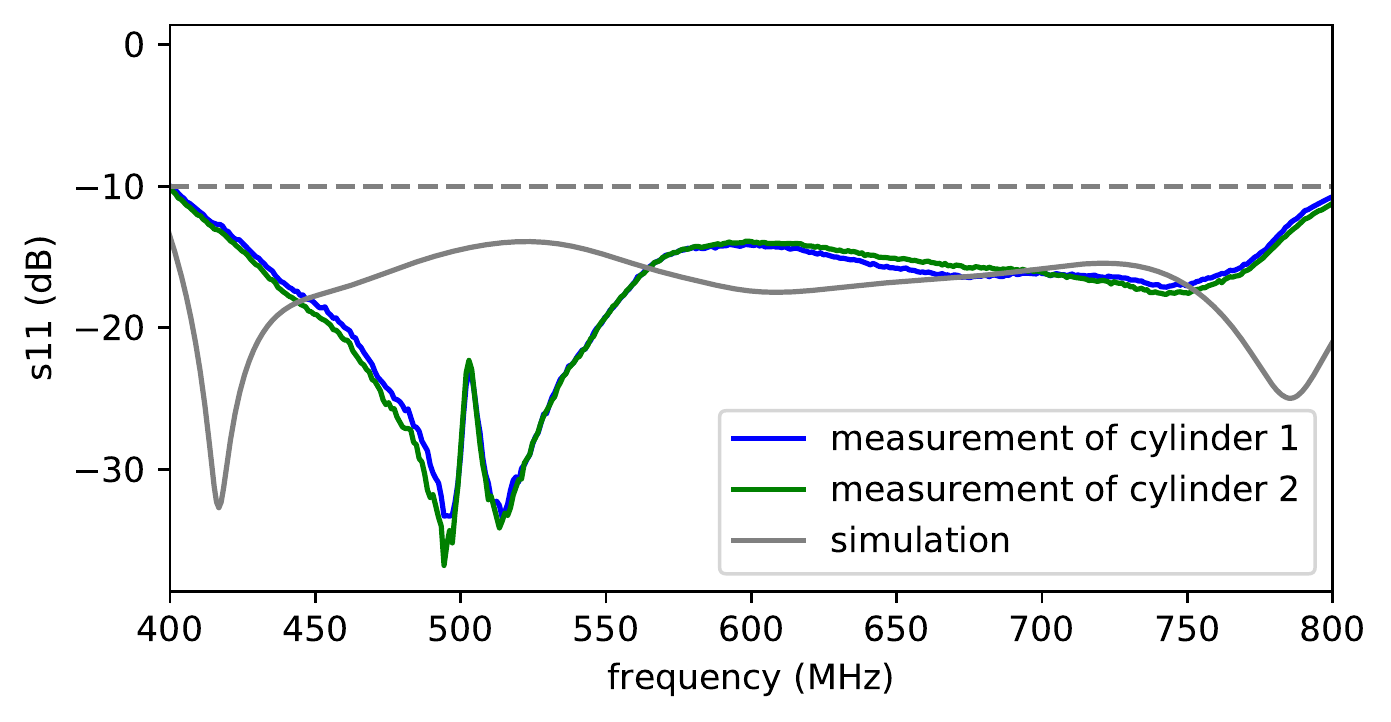}
\end{center}
\caption{$S_{11}$ measurements of the unamplified HIRAX feed in the warm cavities with RF absorber installed. We compare measurement results from the two cylinders that make up the Y-factor measurement system (both at 300\,K), finding they are identical to sub-dB level, and share the same overall level as simulation results, though resonance peak locations differ. These return loss measurements match with published passive feed  measurements of CHIME, which shares the HIRAX antenna design\cite{deng14}. More details are described in the text.} 
\label{s11measandsim}
\end{figure}

The $S_{11}$ measurements used to verify insulation transparency are shown in Figure~\ref{materialtransparency}. As noted, there is no RF absorber present in the cavity for these measurements, so the $S_{11}$ value should be near 0.0\,dB (indicating a fully reflective system). The median value across the spectrum is 0.3\,dB, which can be attributed to losses in the feed. The lines marked `empty' are measurements of the cavity devoid of any insulating foam. There are strong negative features in the $S_{11}$ spectrum which are not present when RF absorber is added (see Figure~\ref{s11measandsim}), so we attribute these features to destructive interference from standing waves at a variety of characteristic distances inside of the cavity. The foam inserts were constructed over the course of six months, in order of: the cryogenic polyurethane foam (2019/03), the fiberglass insert (2019/06), and the full insulation layer (2019/08). In addition, the empty cavity was measured twice (2019/03 and 2019/06), with the $S_{11}$ spectra changing by up to 0.2\,dB. The typical change between the empty cylinder and the full insulation is $<$0.1\,dB, within the range of fluctuations between the two empty measurements. This indicates that the insulating foam layer has no discernible absorption within measurement errors. 

\subsection{Return loss with RF absorber installed} 
For the second set of verification measurements, we performed a series of $S_{11}$ measurements of the full system (including RF absorber) at ambient temperature to verify: (1) that the system does indeed mimic free space and (2) that the two cavities are sufficiently similar to one another, within the -10\,dB design specification for the antenna. These $S_{11}$ measurements were taken over several months using a passive HIRAX feed.

The $S_{11}$ measured for a single polarization in both cylinders is shown in Figure~\ref{s11measandsim}. Also shown is the simulated feed in free-space for comparison, reproduced from Figure~\ref{simulateds11}. The $S_{11}$ profiles evolved slightly in time, but consistently remained at or below -12dB across the full band, indicating a well-matched antenna viewing a system simulating free space. These measurements matched simulations in overall $S_{11}$ level, but had different resonance locations. Although the measurements do not fully agree with the free-space simulations, they appear consistent with measured free-space values shown for a similar feed built for the CHIME experiment\cite{deng14}, and so we attribute differences in resonance locations to differences between the modelled feed in CST and the as-built feed. Despite their differences, these measurements indicate that the cavities are similar to within specifications, with cavity differences accounting for sub-Kelvin uncertainty across our band. This will be explored in Section \ref{sec:systematics}.

\subsection{Blackbody spectrum comparison}
We can perform an additional check to verify that the RF absorber in the cavity is indeed functioning as a blackbody. The RF absorber should emit thermal radiation, and hence have a blackbody thermal spectrum. This aspect of system performance is verified by installing the unamplified feed in one of the cavities, amplifying the resulting signal with commercial amplifiers of known gain and noise temperature, and measuring the resulting spectrum with a spectrum analyzer\footnote{Measurements are made with an R\&S FSH4 Multi-purpose analyzer}. For a $\sim$300\,K absorber in the frequency range 400-800\,MHz, the low-frequency approximation to the blackbody spectrum is valid, providing an estimated power of: 

\begin{equation}
P=GkT\Delta \nu
\end{equation}
where $G$ is the device gain, $k$ is Boltzmann's constant, $T$ is the sum of the physical temperature and device noise temperature, and $\Delta \nu$ is the bandwidth in Hz. We assume a temperature T containing contributions from the temperature of the RF absorber (300\,K), the estimated feed loss (20\,K), and the noise temperature of the first amplifier in the amplifier chain (determined by the data sheets).

We compared this theoretical power with the measured power in dBm from the spectrum analyzer, using a variety of amplification chains. The amplifiers used in this measurement are commercially obtained from Mini-Circuits \footnote{www.minicircuits.com/}, and have available data sheets reporting gain and noise temperature. These amplifiers were chosen because they had good gain in the HIRAX band and have high enough compression points to ensure the measurements would be linear with input power. Using multiple amplification stages brought the signal well above the noise floor of the spectrum analyzer, and 3\,dB attenuators were placed between the amplifiers to reduce reflections and oscillations in the amplified signal.

\begin{figure}[h]
\begin{center}
\includegraphics[width=0.8\textwidth]{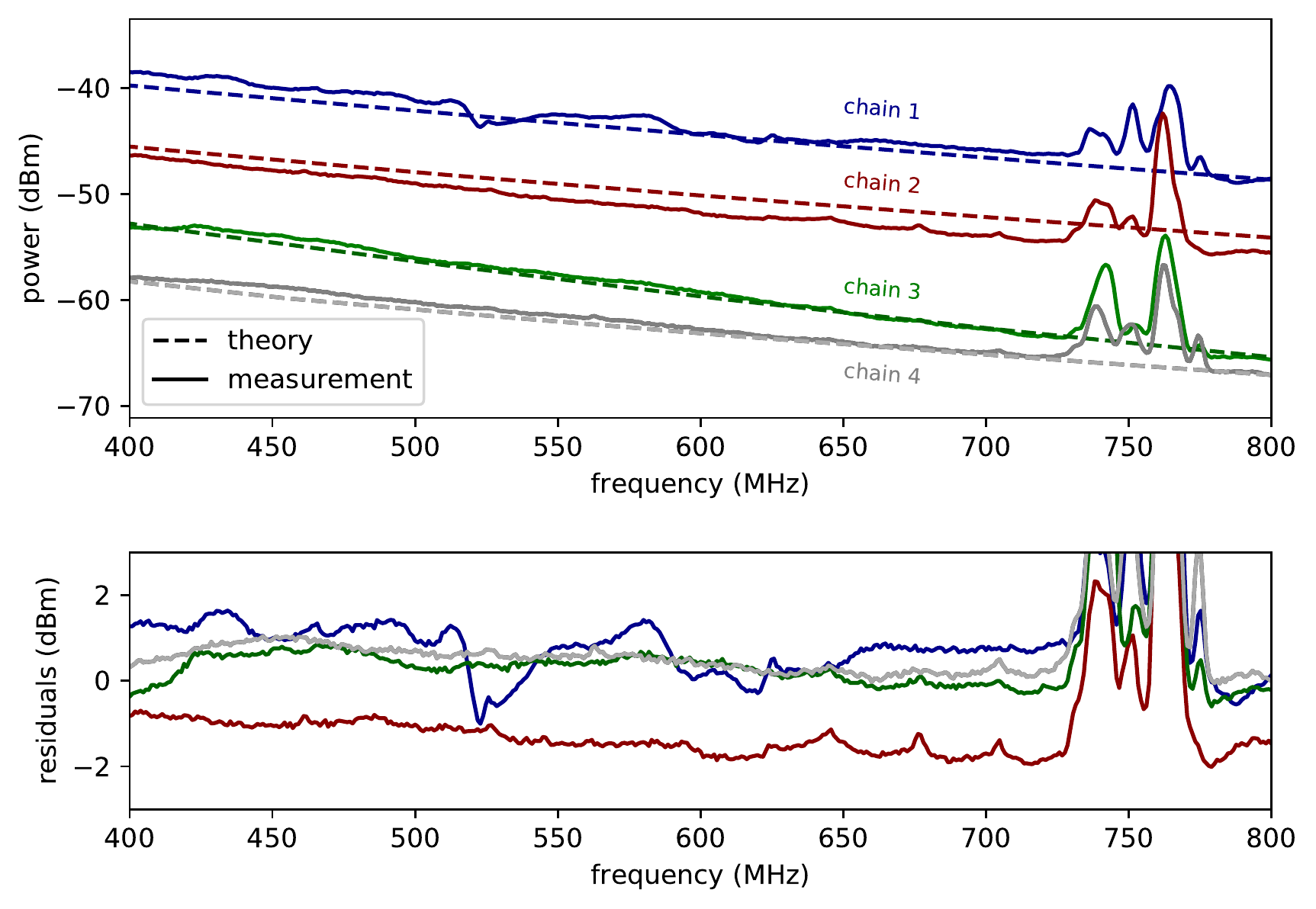}
\end{center}
\caption{Theoretical (blackbody) spectrum, measured spectrum, and residuals for the passive HIRAX feed + four commercial amplifier chains. The top plot shows a comparison between the expected power of the amplifier chains at 300K (dashed lines) and the corresponding spectrum measurements in the experimental system (solid lines). The expected power is computed from P $= G_{\rm chain}(\nu) k_B (T_{\rm load}+T_{\rm loss}+T_{\rm LNA,1}) \Delta \nu$, where $T_{\rm load}$ is the thermal load temperature, $T_{\rm loss}$ is the assumed feed noise temperature from material loss, and $T_{\rm LNA,1}$ is the noise temperature of the first LNA in the chain. The bottom plot shows the residuals, revealing only slight differences between experimental and theoretical values (neglecting the features in the vicinity of 750\,MHz) that could be accounted for by cable loss, gain uncertainty, and other systematics, verifying that we measure a blackbody. The chains are comprised of the following Mini-circuits amplifiers: 
\textbf{chain 1} = ZFL-1000H+ $\to$ ZX60-P103LN+ $\to$ ZX60-P103LN+; 
\textbf{chain 2} = ZX60-112LN+ $\to$ ZX60-P103LN+ $\to$ ZX60-P103LN+; 
\textbf{chain 3} = ZX60-P103LN+ $\to$ ZX60-P103LN+ $\to$ ZX60-P103LN+; 
\textbf{chain 4} = ZX60-P103LN+ $\to$ ZX60-P105LN+ $\to$ ZX60-P103LN+. 
All amplifiers have frequency dependent gain, and all chains include 9dB attenuation.}
\label{ktdeltanu}
\end{figure}

The results comparing the inferred power to the expected blackbody spectrum are shown in Figure~\ref{ktdeltanu}. The different amplifier chains agree with the expected spectrum to within 2\,dBm, which is consistent with contributions we have not taken into account, such as estimated losses from the cable ($<$1\,dB) and systematic errors in the absolute power measurement from the spectrum analyzer ($<$1\,dBm, typ. $<$0.5\,dBm). The additional features in the spectrum between 725-775\,MHz occur near communication bands, which is evidence that we have not eliminated RFI in these measurements using the closed cylinder as the only RFI protection (further RFI mitigation will be employed in future measurements).

\section{Systematics and Uncertainties}
\label{sec:systematics}
In this section we describe and quantify the contributions of statistical and systematic errors to the noise temperature measurement error budget.

\subsection{Statistical uncertainties} 
The noise temperature measurements are limited in integration time, as the antennas under test can cool while attached to the cold cylinder, thereby changing the noise temperature of the LNAs (LNA noise is temperature dependent and lower at lower temperatures). For initial data taking, we averaged 50 samples with sweep time 0.02\,s for a total integration time of 1\,s. This averaging takes $<$10\,s, which is more than adequate to keep the feed from cooling (temperature effects still to be characterized). For an integration over 50 samples in 3\,MHz frequency bins, statistical fluctuations in the individual frequency bins are limited to $\pm 0.1$\,dBm. These fluctuations can be extrapolated to an error in noise temperature by standard error propagation for the Y-factor linear calculation, summing errors in measured $P_{\rm hot}$ and $P_{\rm cold}$ in quadrature, to obtain $\pm$ 4.82\,K for a 30\,K assumed noise temperature. This noise can be smoothed through further binning, and fluctuations in noise temperature can be integrated down between successive noise temperature measurements to further reduce noise. We also observe slight fluctuations in the average spectrum level to within $\pm 0.01$\,dBm over the course of a one-hour measurement, corresponding to a $\pm$ 0.48\,K uncertainty.

\subsection{Contribution of reflections to uncertainties} 

A variety of systematics must be considered for our measurements, including reflections from various system components such as the liquid nitrogen and zotefoam insulation lid. We can use passive feed $S_{11}$ measurements to bound how cavity differences will impact the noise temperature measurements. The measured $S_{11}$ contains three contributions: (i) signal reflected within the feed structure, (ii) losses in the feed (i, ii inherent to the feed), and (iii) signal reflected back to the feed within the cavity (which depends on the cavity and its interplay with the feed). Here we use the measured $S_{11}$ without distinguishing between these contributions since they cannot be differentiated within our test setup. We follow the scheme detailed below:

The amount of energy radiated from a load can be expressed as an equivalent brightness temperature\cite{balanis2016},
\begin{equation}
T_B=(1-\Gamma^2)T
\label{brightnesstemp}
\end{equation}
where $\Gamma$ is the load reflection coefficient and $T$ is load temperature. Supposing a noise temperature $T_{\text{noise}}$, the power measured from this load during a Y-factor measurement would be: 

\begin{equation}
P_B = Gk[T_B+T_{\text{noise}}]\Delta \nu  = Gk[(1-\Gamma^2)T+T_{\text{noise}}]\Delta \nu.
\end{equation}
For the purposes of estimating uncertainty, the reflection coefficient $\Gamma$ can be obtained from $S_{11}$ via
\begin{equation}
\Gamma = 10^{S_{11}\text{[dB]}/20}.
\end{equation}
$S_{11}$ measurements of the Y-factor system provide an upper bound on load reflections, as $\Gamma$ includes reflection contributions from the feed itself as well as from the RF absorber. The following set of equations shows the noise temperature computation with and without corrections for the reflections

\begin{align}
\label{eq:modifiednt1}
T^{\text{true}}_{\text{noise}}&=\frac{(1-\Gamma_{\text{hot}}^2)T_{\text{\text{hot}}}-Y (1-\Gamma_{\text{cold}}^2)T_{\text{\text{cold}}}}{Y-1}\\
\label{eq:modifiednt2}
T^{\text{uncorrected}}_{\text{noise}}&=\frac{T_{\text{hot}}-Y T_{\text{cold}}}{Y-1} 
\end{align}
where $T^{\text{true}}_{\text{noise}}$ is the true noise temperature and $T^{\text{uncorrected}}_{\text{noise}}$ is the noise temperature one would naively calculate by applying a Y-factor computation without accounting for reflections. When conducting a noise temperature measurement, Y is measured directly, but for the purposes of error assessment we compute a theoretical Y value from expected noise temperature and measured reflections:
\begin{equation}
Y=\frac{P_{B, \text{hot}}}{P_{B, \text{cold}}} = \frac{(1-\Gamma_{\text{hot}}^2)T_{\text{hot}} + T_{\text{noise}}}{(1-\Gamma_{\text{cold}}^2)T_{\text{cold}} + T_{\text{noise}}}
\end{equation}

We estimate uncertainties from reflections as the difference between the  ``true'' and ``uncorrected'' cases (eqs. \ref{eq:modifiednt1}, \ref{eq:modifiednt2}) for various measurement scenarios. We take $T_{\text{hot}}$ = 300\,K, $T_{\text{cold}}$ = 77\,K, and $T_{\text{noise}}$ = 30\,K, and compute $\Gamma_{\text{hot}}$ and $\Gamma_{\text{cold}}$ directly from $S_{11}$ measurements in the ambient temperature/cryogenic cavities. 

\begin{figure}[ht]
\begin{center}
\includegraphics[width=\textwidth]{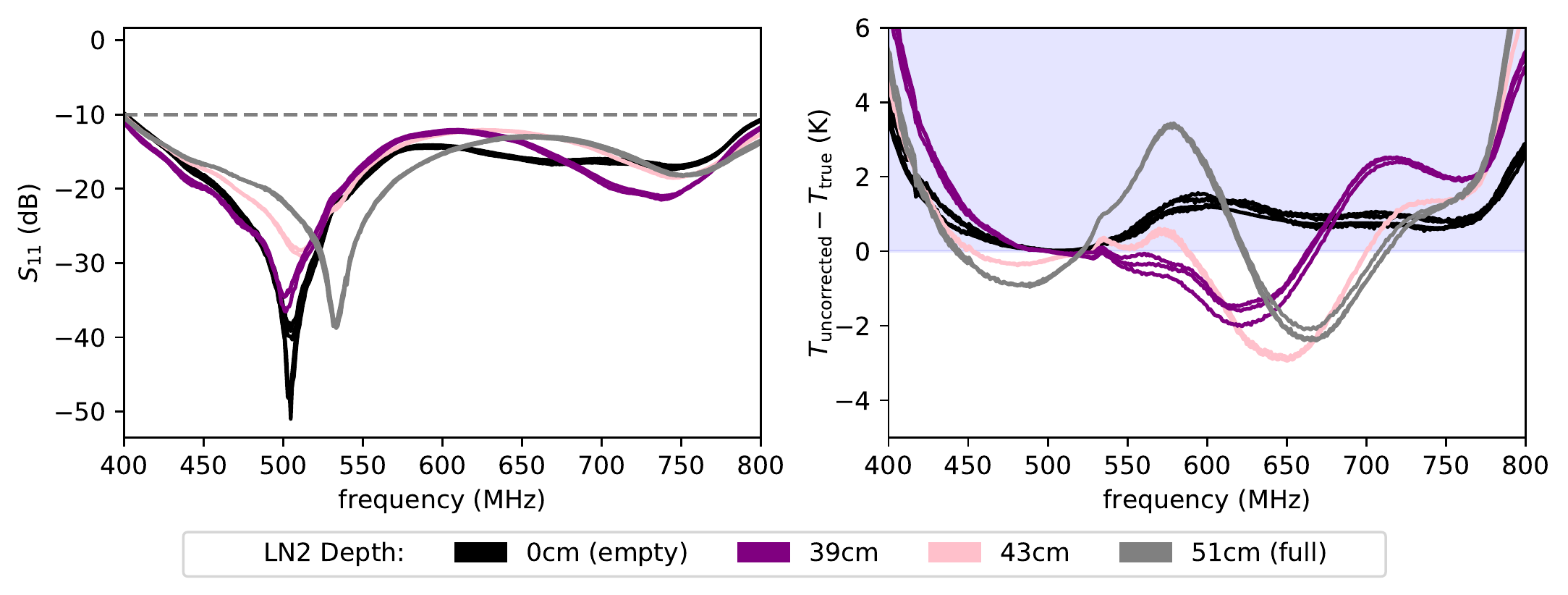}
\end{center}
\caption{$S_{11}$ profiles (as measured by the passive HIRAX feed) at various liquid nitrogen depths and associated uncertainty in noise temperature. (Left): The return loss $S_{11}$ measurements of the cold load exhibit a resonance that shifts in frequency as the cavity fills with liquid nitrogen. Though the resonance shifts, the $S_{11}$ profiles remain below -10\,dB (dashed line) across the full band, which is within the design specification for the feed.
(Right): The uncertainty in noise temperature measurements is computed as described in the main text (equations \ref{brightnesstemp} to \ref{eq:modifiednt2}). The shaded region indicates when this uncertainty provides an upper bound on the measurements, which is the preferred regime. Outside of this regime, the uncertainties remain below approximately 3\,K, which is already a cautiously high bound.}
\label{ln2fillerror}
\end{figure}

Using this description, we estimate the impact of reflections in two different regimes:

\begin{itemize}
    \item When $S_{11} = S_{11, {\rm hot}} = S_{11, {\rm cold}}$, the noise temperature is modified by $T_{\rm uncorrected}=T_{\rm true}/(1-\Gamma^2)$, so $T_{\rm uncorrected}>T_{\rm true}$ (this is the case where the two cylinders are identical and the liquid nitrogen surface is completely sub-dominant). In this regime, we can use the Y-factor measurements without temperature modifications to get an upper bound on noise temperature. 

    \item If $S_{11, {\rm hot}} \neq S_{11, {\rm cold}}$, and they are sufficiently discrepant, we are no longer guaranteed an upper bound on noise temperature. For a noise temperature of 30\,K, and $S_{11, {\rm hot}}$ = -15\,dB, the expected level of the free space HIRAX feed profile, we require $S_{11, {\rm cold}} < -13.5$\,dB to maintain this upper bound. 

\end{itemize}
Figure \ref{ln2fillerror} (right) shows the biases we incur when not accounting for reflections in these two cases. Without liquid nitrogen, we are in the regime where $S_{11, {\rm hot}} = S_{11, {\rm cold}}$, and find the uncertainty at most $\sim$1\,K (black curve). When liquid nitrogen is introduced, there are new reflections ($S_{11, {\rm hot}} \neq S_{11, {\rm cold}}$), giving at most $\pm$ 3.5\,K uncertainty. These uncertainties provide a lower bound at some frequencies and upper bound at others, and remain within 2\,K for much of the frequency band. Although a 2\,K error is within the specification, it is significant enough that removing or accounting for some of the effect of these reflections is desirable.

We measured the $S_{11}$ of the system at various nitrogen depths, as shown in Figure~\ref{ln2fillerror} (left), to assess the reflections off of the liquid nitrogen layer. This measurement revealed that as the distance from the liquid nitrogen surface changes, the location of the `dip' feature, which we associate with constructive interference in the cavity, shifts by $\sim$30\,MHz. The profile continued to change when the cavity was almost full of nitrogen, at a stage when material contractions due to cooling would be complete and RF absorber was nearly submerged, so we suspect this feature is due to reflections from the liquid surface layer and not contraction or altered RF properties from cooling. Still, it is possible that the RF absorber shrinks in cold conditions, and we are currently investigating the impact of perturbations in absorber size in CST.

In addition to reflections from liquid nitrogen, it is important to consider the reflections that might occur off of the insulation lid. This lid is made from zotefoam, which is known for its RF transparent properties. The insulating lid sits directly below the antenna under test, shielding it from the cold nitrogen gas. It is important to understand the lid transparency, as any absorption in the zotefoam could introduce a $>$77\,K component into the 77\,K measurement in a way that is difficult to quantify. To assess the lid RF reflections, we measured the $S_{11}$ of the cold system with and without the zotefoam lid on, finding the measurements identical to $\sim$0.2\,dB for values above -15\,dB, indicating these reflections provide a negligible contribution to the uncertainty.

We also consider the impact that asymmetries in the system might have on reflections, and how that propagates into noise temperature. To assess system symmetry, we took $S_{11}$ measurements at a series of feed rotations (rotating in polarization angle), considering angles 0, 30, 60, and 90 degrees. Comparing these measurements, we see no discernible difference in $S_{11}$ (order $<$0.25\,dB discrepancies for $S_{11}>$-17\,dB), and determine system asymmetries to have negligible contributions to uncertainty. To further support this determination, we also performed Y-factor measurements at the same series of angles, and found no discernible difference in the corresponding noise temperature results. 

\subsection{Estimated contribution of cavity differences to noise temperature calculations} 

Differences between cavities can cause systematic errors in the measurements, and because only one cavity is filled with liquid nitrogen the cavities cannot be interchanged. We can assess the magnitude of these systematic errors through a series of tests at both ambient and cryogenic temperatures. 

\begin{figure}[ht]
\begin{center}
\includegraphics[width=0.9\textwidth]{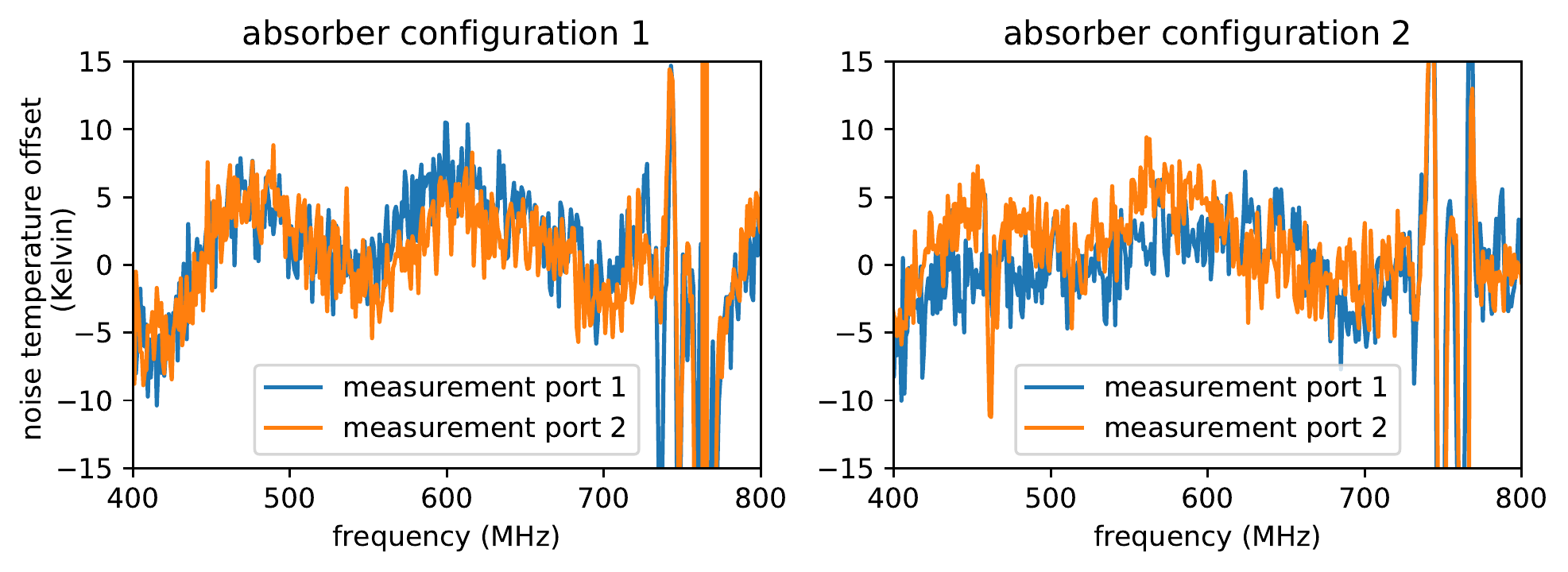}
\end{center}
\caption{Deviations from the expected 30\,K noise temperature due to inherent differences between the two loads (measured at 300\,K). The frequency dependent feature is common across both polarizations and different feeds, and remains when absorber configurations are switched between the two systems. This offset can be removed from the final measurement results. }

\label{simnt}
\end{figure}

For the first of the analyses, we use a HIRAX feed to measure spectra from both cavities when warm, and apply a -4.89\,dB offset to the measurement from the cylinder that will be filled with nitrogen. This result corresponds to the level we would expect from a cold spectrum measurement, assuming a 30\,K noise temperature. We take the warm cavity result as $P_{\rm hot}$ and the offset result as $P_{\rm cold}$ and use the Y-factor method with $T_{\rm hot}$ = 300\,K, $T_{\rm cold}$ = 77\,K to compute a noise temperature. The resulting `mock' noise temperature should have a mean value of 30\,K, with spectral features that are generated only by differences between the spectra in the two cylinders. The results are shown in Figure~\ref{simnt}, where the mean noise temperature of 30\,K has been removed, showing only the expected variations from differences between the cylinders. This feature, a few Kelvin in size, indicates a discrepancy between the RF properties of the two cavities. It is consistent across polarizations and repeatable across feeds and measurement days. It dampens slightly, though remains, when RF absorber is swapped between two systems (demonstrated in Figure \ref{simnt}). These results suggest there could be slight geometric differences in cylinder construction or a difference in insulation transparency (one cylinder has different insulation than the other, for ease of construction).

\begin{figure}[ht]
\begin{center}
\includegraphics[width=0.9\textwidth]{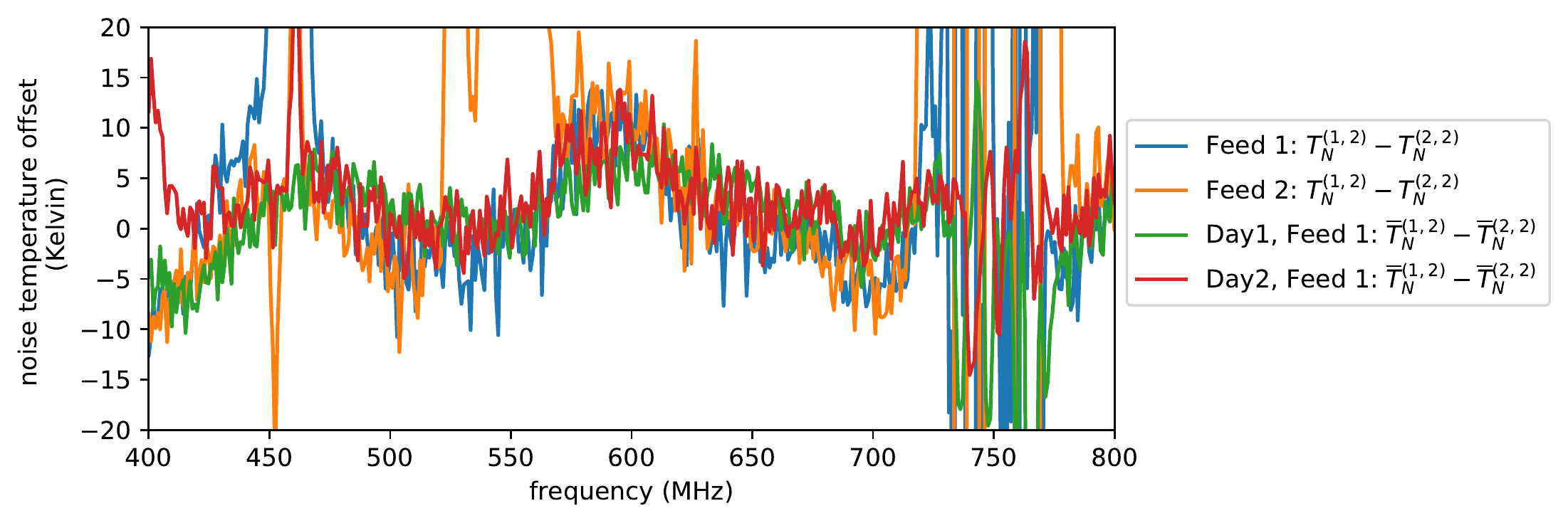}
\end{center}
\caption{ \label{boxcompare}
Characteristic offset in noise temperature measurements due to discrepancies between the two RF cavities, as predicted by warm verification measurements and verified by Y-factor measurements. Here, $T_N^{(1,2)}$ denotes the noise temperature measured with cavity 1 (at 300\,K) as the hot load/ cavity 2 (at 77\,K) as the cold load. We take $\overline{T}_N^{(1,2)}$ to denote the predicted noise temperature computed from two warm measurements (also shown in Figure \ref{simnt}) using a spectrum measurement in cavity 1 (at 300\,K) as the hot power/ a spectrum measurement in cavity 2 (at 300\,K) minus 4.9\,dBm as the ``cold'' power (the 4.9\,dBm offset is consistent with measuring 77\,K). We plot the difference between the noise temperature measured using two different loads for hot/cold measurements and using the same load for hot/cold measurements, along with the difference in predicted noise temperature for two different loads and the same load. The recovered feature is highly consistent across different feeds and different measurement days, and as a result can be removed from final measurements.}
\end{figure}

This sinusoidal feature is recovered in real noise temperature data from hot/cold measurements using a cryogenic load. We take a Y-factor measurement using the same cavity for both hot and cold measurements, where the hot measurement is taken just before the liquid nitrogen fill and the cold measurement is taken directly after. Subtracting this measurement from a measurement taken using the two different cavities at two different temperatures removes any spectral features common to both measurements (e.g. from the noise temperature) and yields features that result from cavity differences. This subtraction is shown in Figure \ref{boxcompare} and reveals the same frequency-dependent profile seen in the warm measurement comparison detailed above. Their strong agreement indicates that the systematics from cavity differences can be subtracted out from measurements to obtain the final noise temperature results.

\subsection{Additional sources of error}

We anticipate additional sources of error in our measurements that we are working to mitigate. These include amplifier drifts, feed temperature fluctuations, cable effects, airflow over the nitrogen surface, and RFI. The full error budget is shown in Table \ref{tab:systematics}.

We have observed long timescale drifts and fluctuations in amplifier gain. Over the course of several months, the amplifier gain appears to drift up to a few dB, presumably in accordance with temperature and humidity conditions in the lab environment. Although we plan to measure noise temperature on far shorter time scales (typically no more than 5 minutes between a hot and cold measurement), these observations highlight the importance of documenting ambient temperature and humidity during the measurements in case absolute power measurements must to be compared across longer time scales at a later point. 

The long-term gain drifts suggest that lab conditions like temperature and humidity impact antenna gain. These effects are critical to consider in the measurement scheme, as we cannot completely isolate the antenna from the cool nitrogen gas leaking out of the insulation when making a cold measurement, and therefore cannot guarantee it measures both loads at the same physical temperature. We have a preliminary data set to characterize this effect, where we continuously measured the spectrum from each load for 10 minutes (the largest timescale we'd consider for one measurement, and ample time for the feed temperature to equilibrate), and then compared the relative drift of spectrum medians in that time frame. Although these measurements suggest no discernible drift, if further analysis indicates this is problematic we can measure the cold load for multiple, shorter time periods. We seek to further mitigate temperature and humidity effects by enclosing the front of the antenna in a zotefoam box through which we flow room-temperature nitrogen gas. This precaution will help keep the antenna at 300K for the full measurement and reduce frost build up. 

In addition to long-term gain drifts, we notice that the HIRAX amplifiers and the spectrum analyzer both take up to an hour to warm up and stabilize in gain and readout when first plugged in. These effects are mitigated by keeping all amplifiers under test powered for at least one hour prior to measurements. 

We also consider the effects of using lossy cables and measuring too close to the amplifier noise floor. The noise temperature of our set up propagates as,
\begin{equation}
T_{\text{noise}} = T_{\text{LNA}} + \frac{T_{\text{cable}}}{G_{\text{LNA}}}+ \frac{T_{\text{LNA, 2}}}{G_{\text{LNA}}G_{\text{cable}}}+ ...
\end{equation}
(following noise of cascaded devices treatment from \textit{Microwave Engineering} \cite{pozar}). From this equation, we see that noise temperature contributions from cable loss should be negligible in the measurement, as the cable sits behind $G_{\text{LNA}}$ = 40\,dB of amplification in the signal chain. Assuming cautiously that $T_{\text{cable}}$ = 300\,K, $G_{\text{cable}} <$ -3\,dB and $T_{\text{LNA,2}} <$ 100\,K, cable loss and second-stage amplifier effects would contribute $<$0.05\,K to the final result. Similarly, measuring near the spectrum analyzer noise floor is not of concern. While measuring close to the noise floor will bias a noise temperature measurement, this is avoidable by adding appropriate second-stage LNAs into the measurement chain.

Previous experiments have found that rapid airflow over a nitrogen surface can artificially raise the nitrogen temperature by 2 degrees\cite{ln2flow}. This was initially a concern for this experiment-- for safety reasons, we draw the nitrogen gas expelled from the tank into the laboratory's HVAC system, and were worried about drawing O2 through the tank in the process. However, this is very likely not an issue in the set up, as we are careful to only remove the gas once it has already been vented, and the zotefoam insulation lid forms a seal that prevents rapid O2 draw. As a worst case, if we were to mistake $T_{\text{cold}}$ as off by 2K, we would see a 3.5K error in the noise temperature. This error is more than we can tolerate, given existing sources of uncertainty from the nitrogen surface and cavity discrepancies.

\begin{table}[h!]
\centering
 \begin{tabular}{||l | l | l||} 
 \hline
 Category & Source of Error & Projected Uncertainty\\ 
 \hline\hline
statistical uncertainty  & spectrum analyzer fluctuations & $<$5\,K*\\
 & spectrum analyzer \& amplifier gain drift  & $<$0.5\,K\\
\hline
 & liquid nitrogen surface reflections & $<$3.5\,K* \\
characterized together & ln2 fill cavity contractions & $\ll$3.5\,K\\
 & insulation reflections & $\ll$3.5\,K\\
\hline
 & cavity RF differences & $<$8\,K*\\
 & airflow over ln2 surface & $\ll$3.5\,K\\
characterized separately & cavity asymmetry & $<$0.1\,K\\
 & cable effects & $<$0.1\,K\\
 & amplifier compression & $<$0.1\,K\\
 \hline
 & temperature/humidity effects & --- \\
characterization in progress & RFI &  ---\\
 & back lobe size & ---\\

\hline\hline

 \end{tabular}
 \caption{ \label{tab:systematics}
Experimental uncertainties and systematics. The main contributors to the error budget are spectrum analyzer fluctuations, nitrogen surface reflections, and cavity differences. These have been marked with a * and are removable or able to be averaged down, as discussed in the text. Additional sources of error are under investigation.}
\end{table}

\section{Preliminary results}
We have taken Y-factor measurement data for three different HIRAX feeds over the course of three separate measurement trials. The procedure is as follows. We first mount the antenna under test on a 12\,in\,$\times$\,12\,in ground plate that slots into the cavity lid, and power the antenna through a bias-t along an SMA cable. We then bolt the mounting plate into place on the ``hot'' cylinder lid, so the antenna is directly facing the RF absorber. We initiate data taking on the spectrum analyzer (previously a Rhode and Schwarz FSH4), and leave the antenna passively taking data (integrating with 3\,MHz bandwidth for 1\,s total integration time), saving a file every 10 seconds for ~5 minutes. At this stage, we move the antenna and plate over to the cold cylinder and bolt the plate onto its lid. We leave it passively data taking for 5 minutes. 
This process is repeated three times. 

Though we established that feed orientation does not impact noise temperature results in a measurable way, we keep the orientation consistent for all measurements. We are also careful to ensure that cables are positioned to reduce strain on the feed SMA connector joints. Later, in the analysis, we tag data files that correspond to hot/cold measurements, and use those files to perform a Y-factor computation.

\begin{figure}[ht]
\begin{center}
\includegraphics[width=\textwidth]{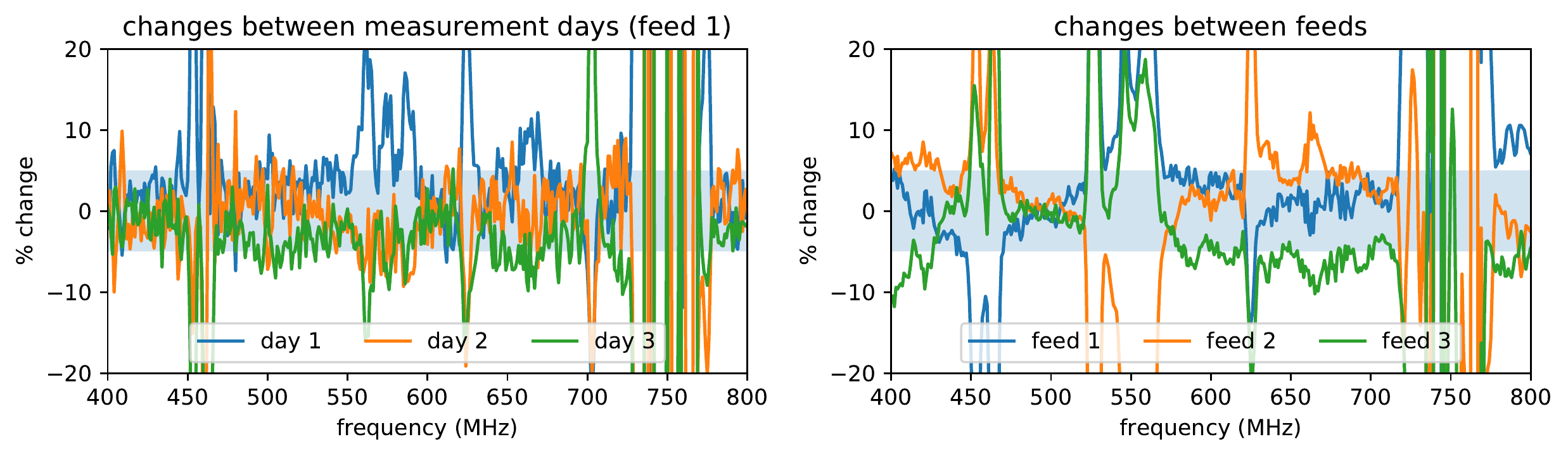}
\end{center}
\caption{Noise temperature consistency for HIRAX feed measurements, shown as a percent change between measurement days (left) and different feeds (right). In both scenarios, excluding RFI, noise temperatures are consistent to within $5\%$ across measurements (light blue band). We note that RFI is persistent across the frequency band.}
\label{hiraxNT}
\end{figure}

Because we have not yet verified noise temperature measurements with known amplifiers, we present relative measurements between different HIRAX feeds and different measurement days instead of absolute noise temperature values. We find the HIRAX feed noise temperature results are repeatable across all three feeds, and three separate measurement days to within $\pm$ 5\% (shown in Figure \ref{hiraxNT}). These preliminary results are encouraging, but further measurements and analysis are required for verification, as outlined in the next section.

\section{Summary and Future Work}
This paper details a system designed and optimized to measure the noise temperature of the HIRAX feeds. In addition, we report results from verification measurements performed to assess systematic and statistical errors. Because HIRAX uses an embedded amplifier in the antenna, a system like this, which can inject an optical signal, is the only way to measure noise temperature in the lab. This measurement will be critical to verifying that the HIRAX feed design meets the noise specifications required to detect the faint cosmological signal of interest. 

This measurement system consists of two identical cavities held at different temperatures to allow for a `Y-factor' measurement. It was designed in the software CST, and constructed over several months in 2019. The construction process included building a cryogenic cavity able to safely and effectively contain over 550\,L of liquid nitrogen. The verification procedure utilized both passive and active HIRAX antennas to measure return loss and spectra of the two cavities and quantify systematic and statistical errors. Initial results indicate that this system is within tolerances, with the main sources of error able to be removed or averaged down (Table \ref{tab:systematics}). 

The verification data sets have shown additional limitations that will need to be accounted for and mitigated. Passive feed measurements indicate that RFI is a contaminant to the data stream, with the potential to significantly bias the noise temperature measurement. To address this challenge we will take two approaches: \textbf{(i)} building a Faraday enclosure around the measurement system and \textbf{(ii)} upgrading the data acquisition from a handheld spectrum analyzer to the ICE board used for HIRAX\cite{bandura2016ice}. The fast sampling rate of the ICE board allows detection of rapid RFI spikes that may currently be averaged into the data. We can then remove contaminated frequency channels in analysis. In addition, the correlator will provide a lower noise floor, and streamline data storage and analysis. 

The current set of verification measurements have focused on quantifying sources of systematic error and assessing whether the cylinders are similar to within specifications. Once these upgrades are completed, we will begin noise temperature verification measurements with this system. We will measure the noise temperature of the HIRAX passive feed with commercial amplifiers of known noise temperatures to assess whether we have met required specification (amplifiers will be the set of four used in Figure \ref{ktdeltanu}). The noise temperature of the commercial amplifiers will also be independently verified with a commercial noise figure meter. From these Y-factor measurements, we expect to recover the amplifier noise temperature plus some consistent contribution from the antenna loss, which we estimate at 10-20\,K from simulations. 

An additional consideration is the antenna backlobe size. Any backlobe in this system will be looking at the laboratory ceiling, a $\sim$300\,K source. As a result, additional power is added during the cold measurement equivalent to 300\,K times the ratio of the backlobe to the total beam. Backlobe effects are a significant contributor to noise temperature ($>$5\,K) if the backlobe makes up more than 1.5\% of the total beam, which sets a requirement that the backlobe remain on average below -17\,dB relative to the peak. Simulations have indicated that the HIRAX feed may be in this regime, such that we will need to account for the additional power in the cold temperature computation. 

Once this verification work is completed and systematics are fully characterized, we expect to have an operational Y-factor measurement system, which will provide the first noise temperature measurements of the HIRAX feed. This system will be used to measure all 256 feeds used for the initial HIRAX deployment, as a spot check on production quality and consistency, as well as next generation feed designs to help inform future prototypes.

\acknowledgments 

This work was made possible by the support from the Yale Wright Laboratory and Yale Center for Research Computing staff and administrators, and the Wright Laboratory computing and machine shop resources. In particular, we acknowledge Frank Lopez, Craig Miller, William Tyndall and James Nikkel for their help constructing the experiment and ensuring personnel safety. Some of the supplementary beam measurements were made in the North Carolina State University Neofabrication Facility's anechoic chamber. The initial feed simulation was generously shared by Andre Johnson, who did some of the feed simulation work for CHIME. This work was supported by a NASA Space Technology Research Fellowship, and is based upon work supported by the National Science Foundation under Grant No. 1751763. KM acknowledges support from the National Research Foundation of South Africa. HIRAX is funded by the NRF of South Africa.

\bibliography{report} 
\bibliographystyle{spiebib} 

\end{document}